# Silicon ring resonator based wavelength conversion via FWM at 10 Gb/s for differential phase-shift keyed signals

F. Li,[1] M. Pelusi,[1] D-X. Xu,[2] R. Ma,[2] S. Janz,[2] B.J. Eggleton,[1,3] and D.J. Moss[1]

[1]*Centre for Ultrahigh-bandwidth Devices for Optical Systems (CUDOS), Institute of Photonics and Optical Science (IPOS), School of Physics, The University of Sydney, NSW 2006, Australia*
[2]*Institute for Microstructural Sciences, National Research Council (NRC-CNRC), Ottawa, ON, Canada K1A-0R6*
dmoss@physics.usyd.edu.au

**Abstract:** We demonstrate all-optical wavelength conversion at 10 Gb/s for differential phase-shift keyed (DPSK) data signals in the C-band, based on four-wave mixing (FWM) in a silicon ring resonator. Error-free operation with a system penalty of ~ 4.1 dB at $10^{-9}$ BER is achieved.

## References

1. *Nature Photonics* Workshop on the Future of Optical Communications; Tokyo, Japan, Oct. 2007. www.nature.com/nphoton/supplements/techconference2007
2. Metcalfe, B., *Toward Terabit Ethernet* Plenary Talk, Optical Fiber Communications 2008, see www.ofcnfoec.org/conference_program/Plenary-video.aspx and www.lightreading.com/tv/tv_popup.asp?doc_id=146223
3. M. Galili, J.Xu, H. C. Mulvad, L.K. Oxenløwe, A. T. Clausen, P. Jeppesen, B.Luther-Davies, S.J. Madden, A.Rode, D.-Y. Choi, M. Pelusi, F.Luan, and B. J. Eggleton, "Breakthrough switching speed with an all-optical chalcogenide glass chip: 640 Gbit/s demultiplexing," Opt. Express **17**, 2182-2187 2009
4. B.J.Eggleton, D.J. Moss, and S.Radic, *Nonlinear Optics in Communications: From Crippling Impairment to Ultrafast Tools* Ch. 20 (Academic Press, Oxford, 2008).
5. H. C. H. Mulvad, M. Galili, L. K. Oxenløwe, H. Hu, A. T. Clausen, J. B. Jensen, C. Peucheret, and P. Jeppesen, "Demonstration of 5.1 Tbit/s data capacity on a single-wavelength channel," Opt. Express **18**(2), 1438-1443 (2010).
6. D. Hillerkuss, R. Schmogrow, T. Schellinger, M. Jordan, M. Winter, G. Huber, T. Vallaitis, R. Bonk, P. Kleinow, F. Frey, M. Roeger, S. Koenig, A. Ludwig, A. Marculescu, J. Li, M. Hoh, M. Dreschmann, J. Meyer, S. Ben Ezra, N. Narkiss, B. Nebendahl, F. Parmigiani, P. Petropoulos, B. Resan, A. Oehler, K. Weingarten, T. Ellermeyer, J. Lutz, M. Moeller, M. Huebner, J. Becker, C. Koos, W. Freude and J. Leuthold, "26 Tbit s$^{-1}$ line-rate super-channel transmission utilizing all-optical fast Fourier transform processing," Nat. Photonics **5**, 364-371 (2011).
7. S. J. B. Yoo, "Wavelength conversion technologies for WDM network applications," J. Lightwave Technol. **14**, 955-966 (1996).
8. B. Ramamurthy and B. Mukherjee, "Wavelength conversion in WDM networking," IEEE J. Sel. Areas Commun. **16**, 1061-1073 (1998).
9. D. Nesset; D. D. Marcenac, P. L. Mason, A. E. Kelly, S. Bouchoule, and E. Lach, "Simultaneous wavelength conversion of two 40 Gbit/s channels using four-wave mixing in a semiconductor optical amplifier," Electron. Lett. **34**, 107-108 (1998).
10. K. K. Chow, C. Shu, L. Chinlon, and A. Bjarklev, "Polarization-insensitive widely tunable wavelength converter based on four-wave mixing in a dispersion-flattened nonlinear photonic crystal fiber," Photon. Technol. Lett. **17**, 624-626 (2005).
11. M. R. E. Lamont, V. G. Ta'eed, M. A. F. Roelens, D. J. Moss, B. J. Eggleton, D. Choy, S. Madden, and B. Luther-Davies, "Error-free wavelength conversion via cross phase modulation in 5 cm of $As_2S_3$ chalcogenide glass rib waveguide," Electron. Lett. **43,** 945 (2007).
12. D. Yeom, E. C. Mägi, M. R. E. Lamont, M. A. F. Roelens, L. Fu, and B.J. Eggleton, "Low-threshold supercontinuum generation in highly nonlinear chalcogenide nanowires," Opt. Letters **33**, 660 (2008).
13. R. Salem, M. A. Foster, A. C. Turner, D. F. Geraghty, M. Lipson, A. L. Gaeta, "Signal regeneration using low-power four-wave mixing on silicon chip," Nature Photonics **2,** 35-38 (2008).
14. B. G. Lee, A. Biberman, A. C. Turner-Foster, M. A. Foster, M. Lipson, A. L. Gaeta, and K. Bergman, "Demonstration of broadband wavelength conversion at 40 Gb/s in silicon waveguides," Photonics Technology Letters **21**, 182-184 (2009).
15. F. Li, M. Pelusi, A. Densmore, R. Ma, D-X. Xu, S. Janz, and D.J. Moss, "Error-free all-optical demultiplexing at 160Gb/s via FWM in a silicon nanowire," Opt. Express **18**, 3905-3910 (2010).
16. H. Ji, M. Pu, H. Hu, M. Galili, L.K. Oxenlowe, K. Yvind, J.M. Hvam, P. Jeppensen, "Optical Waveform Sampling and Error-Free Demultiplexing of 1.28 Tb/s Serial Data in a Nano engineered Silicon Waveguide," IEEE Journal of Lightwave Technology **29**, 426-431 (2011).

17. M.D. Pelusi, F. Luan, S.J. Madden, D.-Y. Choi, D.A. Bulla, B. Luther-Davies, and B.J. Eggleton, "Wavelength Conversion of High-Speed Phase and Intensity Modulated Signals Using a Highly Nonlinear Chalcogenide Glass Chip," Photonics Technology Letters, IEEE **22** (1), pp.3-5 (2010).
18. V.G. Ta'eed, M.R.E. Lamont, D.J. Moss, B. Luther-Davies, S.J. Madden, and B.J. Eggleton, "All Optical Wavelength Conversion via Cross Phase Modulation in Chalcogenide Glass Rib Waveguides", Optics Express **14** 11242 (2006).
19. K. Ikeda, R.E. Saperstein, N. Alic, and Y. Fainman, "Thermal and Kerr nonlinear properties of plasma-deposited silicon nitride/silicon dioxide waveguides", Opt. Express **16**(17), 12987-12994 (2008).
20. D. T.H. Tan, P.C. Sun, Y. Fainman, "Monolithic nonlinear pulse compressor on a silicon chip", Nature Communications **1**, Article Number 116 (2010).
21. A. Gondarenko, J.S. Levy, M. Lipson, "High confinement micron-scale silicon nitride high Q ring resonator Optics Express **17**(14), 11366-11370 (2009).
22. D. Duchesne, M. Ferrera, L. Razzari, R. Morandotti, S.Chu, B.Little, and D. J. Moss, "Efficient self-phase modulation in low loss, high index doped silica glass integrated waveguides", Opt. Exp **17** 1865 (2009).
23. A. Pasquazi, M. Peccianti, M. Lamont, R. Morandotti, B.E Little, S.T. Chu and D.J Moss, "Efficient wavelength conversion and net parametric gain via Four Wave Mixing in a high index doped silica waveguide", Optics Express **18**, (8) 7634-7641 (2010).
24. M. Peccianti, M. Ferrera, D. Duchesne, L. Razzari, R. Morandotti, B.E Little, S. Chu and D.J Moss, "Subpicosecond optical pulse compression via an integrated nonlinear chirper", Optics Express **18**, (8) 7625-7633 (2010).
25. A. Pasquazi, M. Peccianti, Y. Park, B. E. Little, S. T. Chu, R. Morandotti, J. Azaña, and D. J. Moss, "Sub-picosecond phase-sensitive optical pulse characterization on a chip", Nat. Photonics **5** (9) 618 (2011).
26. P. P. Absil, J. V. Hryniewicz, B. E. Little, P. S. Cho, R. A. Wilson, L. G. Joneckis, and P.-T. Ho, "Wavelength conversion in GaAs micro-ring resonators," Opt. Lett. **25**, 554-556 (2000).
27. A. C. Turner, M. A. Foster, A. L. Gaeta, and M. Lipson, "Ultra-low power parametric frequency conversion in a silicon microring resonator," Opt. Express **16**(7), 4881–4887 (2008).
28. M. Ferrera, L. Razzari, D. Duchesne, R. Morandotti, Z. Yang, M. Liscidini, J. E. Sipe, S. Chu, B. E. Little, and D. J. Moss, "Low-power continuous-wave nonlinear optics in doped silica glass integrated waveguide structures," Nat. Photonics **2**(12), 737–740 (2008).
29. L. Razzari, D. Duchesne, M. Ferrera, R. Morandotti, S. Chu, B. E. Little, and D. J. Moss, "CMOS-compatible integrated optical hyper-parametric oscillator," Nat. Photonics **4**(1), 41 (2010).
30. J.S. Levy, A. Gondarenko, M. A. Foster, A. C. Turner-Foster, A. L. Gaeta, and M. Lipson, " CMOS-compatible multiple-wavelength oscillator for on-chip optical interconnects ," Nat. Photonics **4**(1), 37 (2010).
31. A. Pasquazi, R. Ahmad, M. Rochette, M.l Lamont, B.E. Little, S. T. Chu, R. Morandotti, and D. J. Moss, "All-optical wavelength conversion in an integrated ring resonator," Opt. Express **18**, 3858-3863 (2010).
32. F. Morichetti, A. Canciamilla, C. Ferrari, A. Samarelli, M. Sorel, and A. Melloni, "Travelling-wave resonant four-wave mixing breaks the limits of cavity-enhanced all-optical wavelength conversion," Nat. Commun. **2**, 1294- (2011).
33. M. Notomi, A. Shinya, S. Mitsugi, G. Kira, E. Kuramochi, and T. Tanabe, "Optical bistable switching action of Si high-Q photonic-crystal nanocavites, " Opt. Express **13**, 2678-2687 (2005).
34. M. Lamont, L. B. Fu, M. Rochette, D. J. Moss and B. J. Eggleton, "Two Photon Absorption Effects on 2R Optical Regeneration", IEEE Photonics Technology Letters **18,** 1185 (2006).
35. D.J.Moss, L.Fu, I.Littler, B.J.Eggleton, "Ultra-high speed all-optical modulation via two-photon absorption in silicon-on-insulator waveguides", Electronics Letters **41,** 320 (2005).

**1. Introduction**

All-optical nonlinear signal processing is seen [1,2] as a key for future telecommunication networks to overcome the electronic bandwidth bottlenecks as systems evolve towards 640Gb/s [3], Terabit Ethernet [4], and beyond [5,6]. In particular, all optical wavelength conversion is one of the key operations in multi-channel optical communications since it offers the advantages of high-speed and both modulation format and bit rate transparency [7,8]. It has been successfully demonstrated in a number of platforms such as semiconductor optical amplifiers (SOAs) [9] and highly nonlinear silica fiber (HNLF) [10]. However, these platforms all suffer from drawbacks of one form or another, such as free-carrier induced patterning effects in SOAs [9] and fiber dispersion in HNLF [10] that ultimately limit high-speed operation. Therefore, alternative platforms are of significant interest to circumvent these issues and allow the integration of these devices into real systems.

Integrated nanophotonic devices, particularly in highly nonlinear materials, have been a key approach to reducing operating power requirements of all-optical devices by increasing the nonlinear parameter, $\gamma = \omega\, n_2 / c\, A_{eff}$ (where $A_{eff}$ is the waveguide effective area). Both silicon and chalcogenide nanowires have achieved extraordinarily high $\gamma$'s ranging from 15 $W^{-1} m^{-1}$ in ChG waveguides [11] to 95 $W^{-1} m^{-1}$ in ChG tapered fiber nanowires [12], to 300 $W^{-1} m^{-1}$ in Si nanowires [13]. Chip-based all-optical wavelength conversion has been demonstrated in straight waveguides and nanowires in silicon [14-16] and in ChG glass [17,18]. More recently, efficient nonlinear all-optical signal processing has also been demonstrated in glasses having lower nonlinearity, but also much lower linear and nonlinear loss, such as silicon nitride [19-21] and Hydex® glass [22 - 25]. For all of these platforms, an important approach, in addition to increasing $\gamma$, to strengthen the nonlinear interaction between the optical waves in order to increase the device efficiency is through the use of optical resonators [26 -33]. High Q-factor optical resonators have enabled the demonstration of very low power continuous-wave (CW) nonlinear optics [26 - 30], and similar benefits are expected in terms of operation on optical signals containing high bandwidth data. For applications to optical data signal processing, however, the challenge is that the bandwidth of the resonator must be large enough to accommodate all of the spectral components of the optical signal. Indeed, it is only very recently that the first demonstration of optical signal processing based on nonlinear optics in a resonant cavity has been reported [31, 32]. In particular, wavelength conversion at 2.5Gb/s in a single ring [31] and 10 Gb/s in silicon cascaded ring-resonator optical waveguide (CROW) devices [32] has been achieved. However, full system penalty measurements have only yet been reported up to 2.5 Gb/s in these structures [31].

In this paper, we demonstrate all-optical wavelength conversion on a 10Gb/s (33% Return-to-Zero) DPSK pseudorandom bit sequence (PRBS) data stream in the C-band via cavity-enhanced four wave mixing (FWM) in an integrated silicon ring resonator with a Q-factor of ~10,000, a free spectral range (FSR) of ~100 GHz and a full-width at half-maximum (FWHM) of ~20 GHz. We perform bit error ratio measurements and achieve error free operation with a system penalty of ~ 4.1dB at $10^{-9}$ BER using 30 mW of CW pump power. This is the highest bit-rate to date for which all-optical signal processing has been demonstrated with full system penalty measurements based on ring resonator structures, and illustrates that effective enhancement of nonlinear optical processes can be obtained through the use of resonant structures up to these data rates.

**2. Principle of Operation**

The principle of wavelength conversion based on cavity enhanced FWM in ring resonators is well known [27, 28] (Figure 1). A signal and pump beam are launched into the ring resonator from the bus waveguide, producing an idler wave at a frequency shifted from the pump by an amount equal to the signal-pump spacing – a condition resulting from energy conservation. If the signal and pump wavelengths are aligned to a cavity resonance, both beams will be trapped in the resonator and experience significant enhancement in intensity. In addition to enhancing the intra-cavity power, the resonance condition also enhances the effective interaction length with respect to the physical length of the resonator. Further, if the phasematching condition [27, 28] for the FWM process is met (equivalent to the waveguide dispersion being small and anomalous), this results in the idler wave also coinciding with a ring resonance, yielding a triply resonant condition. This has resulted in extremely efficient FWM at CW power levels of only a few mW [27, 28]. Of course for conversion of data and not just CW light, the bandwidth (related to the Q-factor) of the resonator has to be wide enough (how enough Q-factor) to accommodate the signal spectrum. In principle, resonantly enhanced wavelength conversion via FWM in ring resonators can be performed on data any modulation format. However, the performance at a given bit-rate improves significantly for formats having narrower bandwidths (such DPSK used here) than more conventional on-off keying formats such as return-to-zero (RZ) or non-return-to-zero (NRZ) [31].

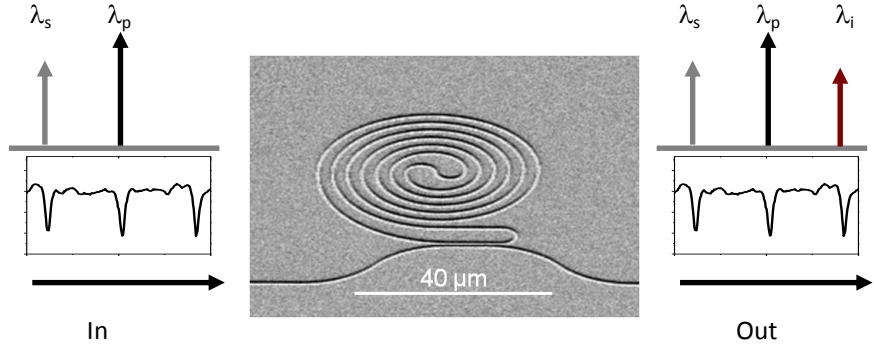

Fig.1 Principle of wavelength conversion based on cavity enhanced FWM in a silicon ring resonator. Center picture shows an SEM of the ring resonator (700 μm length spiral ring, equivalent to about 100 μm radius) while the graphs show the transmission of the ring resonator. $\lambda_p$, $\lambda_s$, $\lambda_i$ refer to the pump, signal and idler wavelengths.

## 3. Experiment

The device is a two-port micro-ring resonator (700 μm length spiral ring) with 5 mm long silicon-on-insulator (SOI) "nanowire" bus. This micro-ring resonator was characterized using an ASE broad-band optical source followed by an optical spectrum analyzer with a resolution of 0.01nm. The experimentally measured transmission function is shown in Figure 3(a), yielding a Q-factor of ~10,000, a free spectral range (FSR) of ~100 GHz and a full-width at half-maximum (FWHM) of ~20GHz. This spectral bandwidth is wide enough to perform the signal processing for 10G RZ-DPSK signal while having a Q-factor that is still high enough to yield significant nonlinear enhancement [28].

The dimensions (and composition) of the nanowire in both the ring and bus sections are the same. The nanowires had a rectangular cross section with dimensions 450 nm wide x 260 nm thick (Figure 2), fabricated by electron-beam lithography and reactive ion etching. The waveguides were coated with SU8, and contained "inverse" taper regions to reduce coupling losses. The measured propagation loss was ~ 3 dB/cm for both TE and TM polarizations and coupling losses were 5 dB/facet for TM and TE polarizations, achieved via lensed fiber tapers with nanopositioning stages, and resulting in a total insertion loss of ~ 11.5 dB (for either polarization). The total dispersion (waveguide plus material) is anomalous and roughly constant at ~ +500 ps/nm/km over the C-band for TE and for TM it is normal and varies between ~ -8,000 ps/nm/km and ~ -16,000 ps/nm/km. we therefore used TE polarization for these experiments to ensure phase matching.

The experimental setup for demonstrating all-optical wavelength conversion on a 10Gb/s DPSK signal is shown in Figure 2. The 10 Gb/s RZ-DPSK signals were generated from a 10GHz integrated mode-locked laser emitting 1.4 ps wide pulses with spectral widths of 1.8 nm centered on a resonance at $\lambda = 1558.68$ nm. A Mach-Zehnder electro-optic phase modulator encoded DPSK data on the pulses at 10 Gb/s with a $2^{15}-1$ pseudorandom bit sequence (PRBS). An optical delay line (ΔT) synchronized the data stream with the 10GHz pulse train. The DPSK data stream was amplified by an erbium-doped fiber amplifier (EDFA) and then filtered by a 0.2nm BPF. The average power of signal was 10 mW (incident, or ~3 mW in the waveguide). The CW pump beam was also centered on a resonance at 1556.985 nm with an incident average power of 30 mW (or ~9 mW in the waveguide). The (TE polarized) signal and pump were combined with a 50:50 coupler before launching into the waveguide with polarizations aligned via polarization controllers (PC). The wavelength converted signal, in the form of the idler product, was then extracted using a 0.22-nm optical BPF and demodulated by an one bit delay line interferometer before detection with a 40 Gb/s receiver (RX). Note that the destructive port of a DPSK demodulator was used in our experiments. An EDFA was used at the receiver end to improve the optical signal-to-noise ratio (OSNR) of the detected signals. We performed system measurements in order to evaluate the transmission bit error rate (BER) and system penalty of the device.

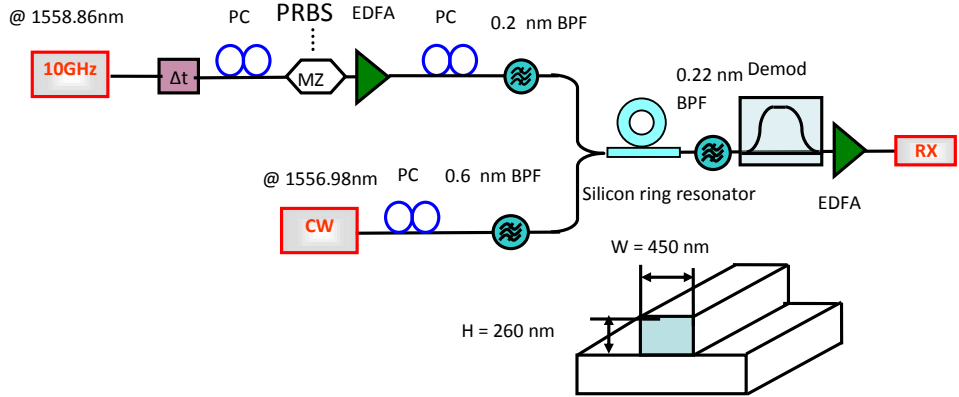

Fig.2 Experimental setup for the wavelength conversion for 10 Gb/s (33% RZ) DPSK signals based on a silicon ring resonator.

## 4. Results and discussion

The optical spectra at the waveguide input is shown in Figure 3 (b). The inset illustrates that the signal spectrum of the 10G RZ-DPSK signal is matched well with the resonance linewidth of the ring. The optical spectra at the waveguide output (Figure 3 (c)) shows the probe and signal spectra as well as the spectrum of the wavelength converted idler at 1555.2 nm generated by FWM, showing a conversion efficiency of ~ -30 dB. A comparison of the output spectra with the signal and pump beams tuned on versus off resonance (0.4 nm shift) clearly shows that the generation of the FWM idler is enhanced by the ring resonance by as much as ~13 dB improvement in signal to noise ratio, critically improving the quality of the generated wavelength conversion signal. This is seen in the eye-diagrams of the idler when tuned on and off the resonance, respectively. Figure 4 shows the demodulated eye diagram of the input DPSK data stream along with the wavelength conversion signal when tuned on and off the resonance of the ring, all extracted by tunable band-pass optical filters followed by an optical sampling oscilloscope. All traces were measured with a 65 GHz detector and sampling oscilloscope. The eye diagram of the 10Gb/s demodulated wavelength converted signal, under resonant conditions, is clearly open, highlighting the effective enhancement of the silicon ring resonator. Finally, the performance of this chip-based wavelength conversion device for both on and off resonant conditions was confirmed with bit error-rate (BER) measurements. Figure 5 shows the system penalty measurements, indicating that absolute (without any forward error correction (FEC)) error-free operation at $10^{-9}$ bit error rate (BER) is achieved, with a system penalty, relative to the un-converted signal transmitted on-resonance through the device (which term "back-to-back" as the reference, following [31]) of ~4.1 dB (at $10^{-9}$ BER) with both the pump and signals tuned on-resonance, whereas with the pump and signals tuned off-resonance, error-free operation could not be obtained – a minimum of only $10^{-6}$ was achievable. This demonstrates the significant impact that the resonant enhancement of the ring has on the nonlinear conversion process – particularly with regard to the BER results.

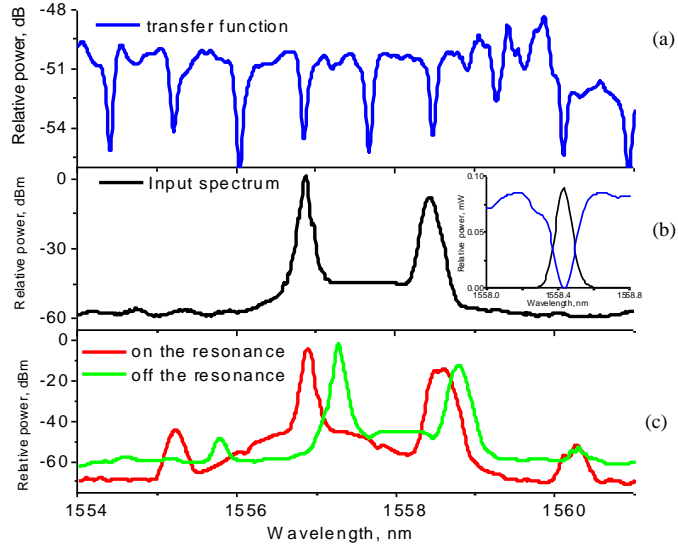

Fig.3 Optical spectrum measured on an OSA: (a) the transfer function of the ring resonator, (b) input spectra of CW pump with 10G signal, Inset: 10G RZ-DPSK signal spectrum with ring resonance (c) output spectra.

With very low input power, we still observed thermo-optical non-linearity [33]. The cavity resonances red shifted by about 0.3 nm when the input pump power was increased from 30 to 40 mW. By fine tuning the pump wavelength, we adjusted the wavelength slightly off-set from the resonance wavelength in order to obtain a stable working point, at the expense of sacrificing a little of the resonant enhancement. The error-free operation of our device clearly demonstrates that silicon ring resonator can effectively reduce the required power for error-free high-speed all-optical signal processing. We note that similar experiments in straight silicon nanowires [14] required ∼ +26dBm of CW pump power – more than 10dB higher than in these experiments. However, a direct comparison is not completely meaningful since many of the parameters in [14] were different, such as the device dimensions, bit rate etc.. In terms of the ultimate speed limits of this device, given the fact that error-free wavelength conversion via FWM in silicon nanowires has been demonstrated at data rates from 40 Gb/s to 1.28 Tb/s [14-16], we anticipate that ring resonators will be capable of operating at higher bit rates by utilizing more advanced modulation formats (to reduce the signal spectral width for a given data baud rate), as well as higher order filter designs (eg., CROW devices) [32] to broaden the device bandwidth. We also anticipate a significant reduction in both the system penalty and operating pump power by reducing coupling and propagation losses as well as optimizing the waveguide dispersion to be closer to the ideal for FWM, namely, small but anomalous, over a broader wavelength range.  Ultimately two-photon absorption effects [34, 35] will limit the device performance but this work shows that despite TPA and the ensuing free carriers, that error free operation can still be achieved.

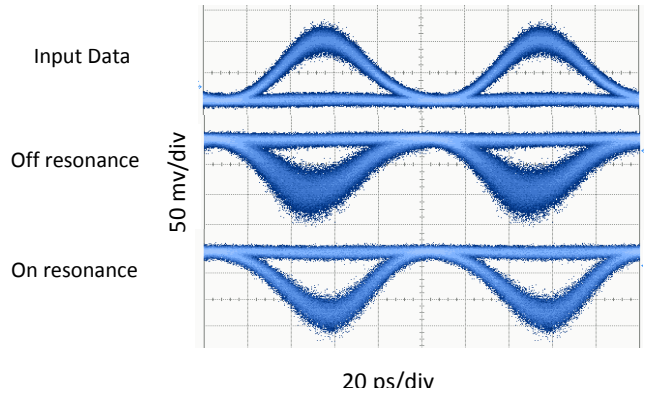

Fig.4 Eye diagrams (b) corresponding to (top) the 10 Gb/s input DPSK signal at λ = 1558.98 nm, (middle) the 10 Gb/s DPSK idler off the resonance at λ = 1555.86 nm and (bottom) on the resonance at λ = 1555.2 nm.

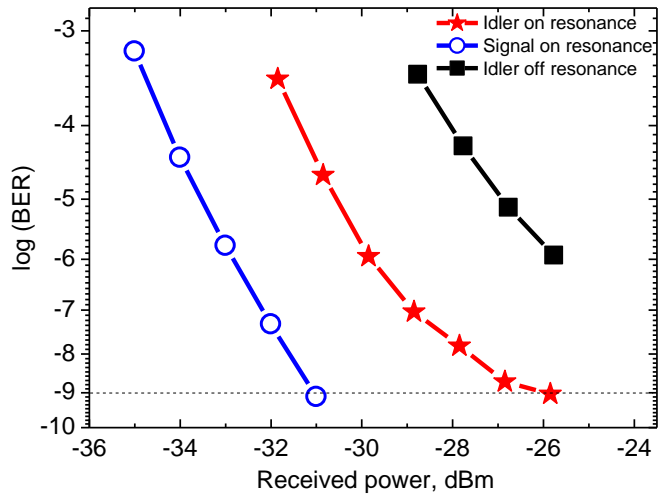

Fig.5 BER measurements for the signal on-resonance (blue lines) and the idler on-resonance (red line) generated by a 9 mW signal and a 30 mW probe. The relative power penalty between the two curves at a BER of $10^{-9}$ is 4.1 dB. The BER measurement for the idler off resonance is shown in black.

## 5. Conclusions

We demonstrate all-optical wavelength conversion on DPSK PRBS data at 10 Gb/s (33% Return-to-Zero) in the C-band via FWM in a silicon ring resonator. We perform bit error rate measurements and achieve error free operation with a penalty of 4.1 dB at $10^{-9}$ BER.

## Acknowledgements

This research was supported by the Australian Research Council (ARC) Centers of Excellence (COE), Discovery Projects and Federation Fellowship programs.